\documentclass[prd,twocolumn,preprintnumbers,superscriptaddress,nofootinbib]{revtex4}
\pdfoutput=1

\usepackage{amsmath}
\usepackage{graphicx}
\usepackage{xspace}
\usepackage{epsfig}
\usepackage{units}
\usepackage{slashed}
\usepackage[hyperfootnotes=false]{hyperref}
\usepackage{gensymb}
\usepackage{cancel}
\usepackage{multirow}
\usepackage[normalem]{ulem}

\def \epsilon {\varepsilon} 

\newcommand{\sw}{\ensuremath{s_W}}
\newcommand{\cw}{\ensuremath{c_W}}
\newcommand{\alr}{\ensuremath{A_{LR}}}
\newcommand{\sstw}{{\ensuremath{\sin^2\theta_W}}}

\newcommand{\msbar}{\ensuremath{\overline{\text{MS}}}\xspace}

\allowdisplaybreaks

\begin{document}

\preprint{ACFI-T19-15}

\title{Parity-Violating M\o ller Scattering at NNLO: Closed Fermion Loops}

\author{Yong Du}
\email{yongdu@umass.edu}
\affiliation{Amherst Center for Fundamental Interactions, Physics Department, University of Massachusetts Amherst, Amherst, MA 01003 USA}

\author{Ayres Freitas}
\email{afreitas@pitt.edu}
\affiliation{Pittsburgh Particle-physics Astro-physics \& Cosmology Center
(PITT-PACC),\\ Department of Physics \& Astronomy, University of Pittsburgh,
Pittsburgh, PA 15260, USA}

\author{Hiren H. Patel}
\email{hpatel6@ucsc.edu}
\affiliation{Department of Physics and Santa Cruz Institute for Particle Physics,
University of California, Santa Cruz, CA 95064, USA}

\author{Michael J. Ramsey-Musolf}
\email{mjrm@physics.umass.edu}
\affiliation{Tsung-Dao Lee Institute, and School of Physics and Astronomy, Shanghai Jiao Tong University, Shanghai 200240, China}
\affiliation{Amherst Center for Fundamental Interactions, Physics Department, University of Massachusetts Amherst, Amherst, MA 01003 USA}
\affiliation{Kellogg Radiation Laboratory, California Institute of Technology, Pasadena, CA 91125 USA}

\hypersetup{
    pdftitle={Closed fermion loop contributions to virtual radiative corrections to parity violating asymmetries in polarized electron scattering},
    pdfauthor={YD, AF, Hiren Patel, MJRM}
}

\begin{abstract}
A complete, gauge-invariant computation of two loop virtual corrections involving closed fermion loops
to the polarized M{\o}ller scattering asymmetry is presented. The set of contributions involving two closed fermion loops and the set involving one closed fermion loop are numerically similar in magnitude to the one-loop bosonic corrections and yield an overall correction of 1.3\% relative to the tree-level asymmetry.
We estimate sizes of remaining two-loop contributions and discuss implications for the upcoming MOLLER experiment.
\end{abstract}

\maketitle

\section{Introduction}
Precision measurements of electroweak processes have played a vital role in the development and testing of the Standard Model of particle physics. With the discovery of the Higgs boson at the CERN Large Hadron Collider, the focus of precision tests now falls squarely on the search for signs of physics beyond the Standard Model (BSM). While a variety of open questions clearly point to the existence of BSM physics, it remains to be determined at what mass scale this physics lives and how it interacts with the known elementary particles of the SM. 

A powerful probe in this context is parity-violating electron scattering (PVES).  The relevant observable in PVES experiments is the asymmetry $\alr$  in the cross-section when otherwise identical beams of longitudinally-polarized electrons of left (L) and right (R) helicities scatter from a fixed target
\begin{equation}
A_{LR} = \frac{d\sigma_L - d\sigma_R}{d\sigma_L + d\sigma_R}\,.
\label{eq:alr2}
\end{equation}
Historically, the measurement $\alr$ in deep-inelastic electron-deuteron scattering singled-out the Glashow-Weinberg-Salam theory \cite{Glashow:1961tr,Salam:1968rm,Weinberg:1967tq} of the electroweak interaction from other alternatives and provided the first measurement of the all-important weak mixing angle, $\theta_W$.  Improved results were later obtained by a variety of PVES measurements at low energies, along with observations of parity violation in atomic Cesium and neutrino-nucleus deep-inelastic scattering.  Parity-violating (PV)  M\o ller scattering provides one of the theoretically cleanest such tests. The first measurement of this asymmetry was made by the E158 Collaboration at SLAC in the mid-2000's \cite{Anthony:2005pm}, yielding a confirmation of the predicted running of $\sstw$ with $6 \sigma$ significance.

A new, more precise measurement of the PV M\o ller asymmetry--- 
dubbed MOLLER and approved to run at the Jefferson Lab \cite{Mammei:2012ph,Benesch:2014bas}---aims to determine $\alr$ with 2.4\% uncertainty. Assuming only SM contributions, the MOLLER experiment will yield a value of $\sstw$ with an uncertainty comparable to the earlier determinations in high energy $e^+e^-$ annihilation.  Within the Standard Model, this measurement can be interpreted as a precision test of the scale-dependence of $\sstw$ \cite{Czarnecki:1998xc,Erler:2004in,Erler:2017knj}. Its value at $\mu=m_Z$ can be obtained either from fits to high energy electroweak precision observables, while PVES experiments yield $\sstw$ at a low scale $\mu \ll m_Z$.

More significantly, MOLLER will provide a new probe for BSM physics that could reside at either high or low-mass scales. Examples include $1-10$ TeV doubly-charged scalar bosons that are implied by left-right symmetric models for the non-vanishing neutrino masses\cite{Dev:2018sel} and a light ``dark'' $Z$ boson that, under certain conditions, may also account for the observed deviation of the muon anomalous magnetic moment from SM predictions\cite{Davoudiasl:2012qa,Davoudiasl:2014kua,Davoudiasl:2015bua}. In both examples, the PV M\o ller asymmetry provides a complementary probe to other tests at low- and high-energies. 

The unique potential of the PV M\o ller scattering follows from two features: the purely leptonic character of the process and a fortuitous suppression of the leading-order (LO) asymmetry by $1-4 \sstw$ ($\sstw$ is numerically close to $1/4$).   Specifically, the theoretical prediction for the PV M\o ller asymmetry can be written as \cite{Derman:1979zc}
\begin{equation}
A_{LR} = \frac{G_\mu Q^2}{\sqrt{2}\pi
\alpha}\,\frac{1-y}{1+y^4+(1-y)^4}(1-4\sin^2\theta_W + \Delta Q_W^e)
\label{eq:alr}
\end{equation}
where $y=Q^2/s$, and $\Delta Q_W^e$ accounts for radiative corrections.

Some terms (SM and possibly BSM) entering through $\Delta Q_W^e$ do not carry the factor $1-4\sstw$ and thus their relative impact is enhanced.  Importantly, the NLO electroweak corrections, whose relative impact should be nominally $\mathcal{O}(\alpha)\sim 0.01$ are roughly 40\% in magnitude\cite{Czarnecki:1995fw}. These corrections are dominated by contributions from closed-fermion loops that enter the running of $\sstw$.  The $WW$ and $\gamma Z$ boxes also produce sizeable corrections.
Given this enhanced NLO sensitivity, it is important to determine the magnitude of NNLO SM corrections if one wishes to interpret correctly a 2.4\% measurement of $\alr$ in terms of BSM physics.  Partial results at the NNLO level have been presented in Refs.~\cite{Aleksejevs:2011de,Aleksejevs:2012xua,Aleksejevs:2015dba,Aleksejevs:2015zya}. Furthermore, second-order QED effects have been studied in the context of electron-proton scattering \cite{Bucoveanu:2019hxz}, which shares many features with electron-electron scattering.

In what follows, we report on a computation of all NNLO contributions involving closed fermion loops. This subset of the complete NNLO electroweak corrections is gauge-invariant and, thus, constitutes a well-defined contribution to the asymmetry. Since closed fermion loops dominate the NLO corrections and since they entail a sum over all colors and flavors of SM fermions, we expect them to generate the leading effect at NNLO. We find a resulting 1.3\% correction to the LO asymmetry, again significantly larger than one might expect based on $\alpha/4\pi$ counting. As we discuss below, we expect the contributions from the remaining NNLO corrections to be smaller in magnitude. We thus anticipate the overall uncertainty in the SM prediction for $\alr$ lies below the planned experimental uncertainty.

\section{Method}

\begin{figure}
\centering
\includegraphics[scale=0.5]{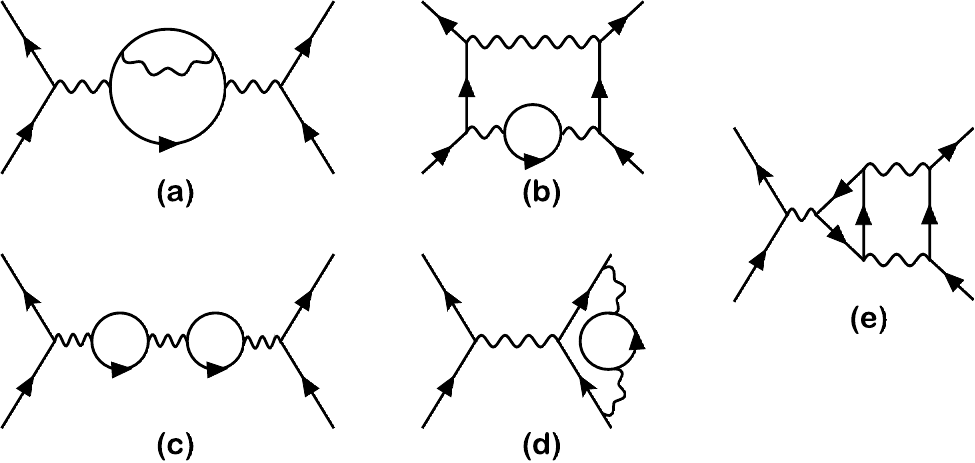}
\caption{Examples of two-loop Feynman diagrams with at least one closed fermion loop.}\label{fig:diaga}
\end{figure}

We calculate the left-right asymmetry by expanding eq.~(\ref{eq:alr2})
up to two-loop order. 
Non vanishing contributions to $\alr$ arise from the interference of a purely electromagnetic amplitude with the PV component of the weak neutral current amplitude arising from $Z$-exchange, with the electromagnetic contributions dominating the denominator in eq.~(\ref{eq:alr2}).  For these building blocks, the NNLO corrections to $d\sigma$ stem from
two-loop matrix elements contracted with Born amplitudes, as well as the
interference of two one-loop matrix elements. The two-loop matrix elements 
receive contributions from genuine two-loop self-energy, vertex and box
diagrams, and from one-particle reducible two-loop diagrams (see
Fig.~\ref{fig:diaga} for examples).

When counting the numbers of fermion loops, we do so at the level of the final
observable $A_{LR}$. This means that we include contributions from two-loop
diagrams with at least one closed fermion loop, as well as products of a
one-loop diagram with fermion loop with another one-loop diagram.  However, for
consistency we exclude products of one-loop diagrams without closed fermion
loops that could arise from interference terms obtained by expanding the
denominator of Eq.~\eqref{eq:alr2} to two-loop order.

Logarithmically enhanced contributions from real photon emission cancel out in the ratio \eqref{eq:alr2}
and thus it is not necessary to compute them. Similarly,
all infrared (IR) singularities in the virtual loop contributions also
cancel. However, this cancellation is only achieved when combining all
terms that contribute to $A_{LR}$, but individual loop diagrams are IR
divergent.  We use a small photon mass $m_\gamma$ and electron mass $m_e$ to regulate the soft and
collinear divergencies appearing at intermediate steps, respectively. 

In addition, there are ultraviolet (UV)
divergences, for which we employ dimensional regularization. The UV divergencies
are eliminated by appropriate renormalization conditions. We employ a
renormalization scheme similar to Ref.~\cite{Czarnecki:1995fw}, where the
on-shell (OS) scheme is used for the electromagnetic coupling and the
$Z$-boson, Higgs boson and fermion masses. For the weak mixing angle $\sin^2
\theta_W$ we use the \msbar renormalization scheme to make contact with
descriptions of $\sin^2\theta_W$ as a running parameter in the literature.
Specifically, we use the \msbar scheme in the full SM, without any decoupled
degrees of freedom, which ensures that $\sin^2
\theta_W(\mu)_{\msbar}$ is gauge invariant.
By default, the scale choice $\mu=m_Z$ is used in the following.
Expressions for the on-shell counterterms can be found in
Ref.~\cite{Freitas:2002ja}.

To guarantee the cancellation of UV divergencies, one must impose the relation
$\sin^2 \theta_W = 1-m_W^2/m_Z^2$, where
$m_W$ and $m_Z$ are the renormalized gauge boson masses in any given renormalization scheme (not necessarily the OS scheme).
This implies that one cannot choose an
independent renormalization condition for $m_W$, but instead the $m_W$
counterterm is restricted to
\begin{equation}
\delta m_W^2 = (1-\sw^2)\delta m^2_{Z,\rm OS} - m_Z^2 \delta s^2_{W,\msbar}\,,
\end{equation}
where $\sw^2 \equiv \sin^2\theta_W(\mu)_{\msbar}$. The renormalized mass,
$m_{W,\rm ren}$, defined in this fashion differs from the OS mass, $m_{W,\rm
OS}$, and an additional finite correction would be needed to relate the two.
However, given that $m_W$ is never used as an input or output in our
calculation, this correction is never explicitly needed in our case.

When performing calculations in dimensional regularization, one has to be
careful about the treatment of $\gamma_5$. In $d\neq 4$ dimensions, the
anticommutation rule $\{\gamma^\mu,\gamma_5\}$ is incompatible with the trace identity
$\text{tr}\{\gamma^\alpha\gamma^\beta\gamma^\gamma\gamma^\delta\gamma_5\} 
= -4i \epsilon^{\alpha\beta\gamma\delta}.$ 
Contributions from such traces arise from vertex
diagrams with a triangle sub-loop, see Fig.~\ref{fig:diaga}~(e), and from box
diagrams. However, in both of these cases, contributions stemming from $\epsilon$-tensors
are UV-finite (after including the sub-loop counterterms for the box graphs) and
thus can be computed in 4 dimensions without ambiguity\footnote{A similar
argument holds for a set of useful identities for 4-fermion scattering matrix
elements \cite{Sirlin:1981pi,Denner:2005fg}.}.

Throughout the calculation, we exploit the hierarchy of scales $m_e^2 \ll Q^2 \ll
m_{\rm weak}^2$, where $m_{\rm weak} \sim m_W,m_Z,\linebreak[0]m_H,m_t$. In
practice, this means that the mass of the external electrons is set to zero
everywhere except where it is needed to regularize collinear singularities.
Furthermore, we perform a large-mass expansion for $m_{\rm weak}^2 \gg Q^2$, up
to order $m_{\rm weak}^{-2}$, which is the leading order needed for parity
violating effects \cite{Derman:1979zc,Czarnecki:1995fw}. This expansion is based
on the method of regions \cite{Beneke:1997zp,Smirnov:1999bza,Smirnov:2002pj}, and in many cases
it leads to products of one-loop integrals and two-loop vacuum integrals, which
are analytically known \cite{tHooft:1978jhc,Beenakker:1988jr,Ford:1992pn,Davydychev:1992mt}. 
The only exception are vertex and box diagrams
with a light-fermion loop inside a photon or photon-Z propagator (see
Fig.~\ref{fig:diaga}~(b)). Here ``light fermion'' refers to any SM fermion except the top quark.

We evaluate these two-loop integrals using the numerical dispersion
integral technique \cite{Bauberger:1994by} (see also
Refs.~\cite{Awramik:2006uz,Aleksejevs:2018tfr}). Since only the transverse part
of the sub-loop self-energy $\Sigma_{\mu\nu}(k^2)$ contributes, we
decompose it as
\begin{align}
\Sigma_{\mu\nu} &= [g_{\mu\nu} k^2 - k_\mu k_\nu] \Pi_{\rm T}(k^2)\,, \\
\Pi_{\rm T}(k^2) &= c_{\epsilon} + \frac{k^2}{\pi} \int_0^\infty d\sigma \; 
 \frac{\text{Im}\{\Pi_{\rm T}(\sigma)\}}{\sigma(\sigma-k^2-i0)}\,.
\end{align}
The contribution of a fermion with mass $m_f$ is given by
\begin{align}
c_{\epsilon} &= \frac{N_cg_1g_2}{12\pi}\biggl(\frac{1}{\epsilon} +
\ln\frac{\mu^2}{m_f^2}\biggr)\,, \\
\text{Im}\{\Pi_{\rm T}(\sigma)\} &= \frac{N_cg_1g_2}{12\pi}
\Bigl(1+\frac{2m_f^2}{\sigma}\Bigr) \notag \\
&\quad \times \sqrt{1-\frac{4 m_f^2}{\sigma}} \;\Theta(\sigma-4 m_f^2)\,,
\end{align}
where $1/\epsilon = 2/(4-d)$, $\Theta(x)$ is the Heaviside step function, and $N_c
= 1$ (3) for leptons (quarks).
The couplings are $g_1g_2=e^2Q_f^2$ and $g_1g_2 =
\frac{e^2Q_f(2\sw^2Q_f-I_{3f})}{2\sw\cw}$ for the photon and photon-Z
self-energy, respectively. Inserting these expressions into the outer loop leads to integrals
of the form
\begin{align}
\begin{aligned}
\int \frac{d^dk}{i\pi^{d/2}} &\; \frac{N(k)}{\prod_i [(k+p_i)^2-m_i^2+i0]} \\
&\hspace{-1ex}\times
\biggl[ c_\epsilon - \frac{1}{\pi}\int \frac{d\sigma}{\sigma} \;
\text{Im}\{\Pi_{\rm T}(\sigma)\}
 \frac{k^2}{k^2-\sigma+i0}\biggr ]. \label{eq:disp2}
\end{aligned}
\end{align}
Here $p_i$ are sets of external momenta, as they appear in a given vertex or box
diagram, and $N(k)$ accounts for dot products ($k^2, k\cdot p_i$) and
$\cancel{k}$ in the numerator. The $k$-integral in Eq.~\eqref{eq:disp2} is a
conventional one-loop integral, which can be performed analytically and reduced to
basic scalar one-loop functions using the standard Passarino-Veltman method. The
remaining $\sigma$ integral, which is UV-finite, is easily evaluated
numerically with high precision. It is interesting to note that the $\sigma$
integrals involving $\log m_\gamma^2$ and
$\log m_e^2$ may be performed analytically so that the cancellation of IR singularities in the full result
can be checked algebraically.

These dispersion integrals are not well-defined for light quarks ($f=u,d,s$) in
the inner loop, since the dominant contribution to the integral arises from
region where $k^2 \sim m_f^2$, where hadronization effects become important. In
fact, the same problem already occurs at the one-loop level for the self-energy
contribution to the $\gamma$-$Z$ self-energy in the $t$- and $u$-channel
\cite{DePorcel:1995nh,Czarnecki:1995fw}, due to the fact that $Q^2 < \Lambda^2_{\rm QCD}$.

The non-perturbative hadronic corrections can be approximately accounted for by
using dressed quark masses. A well-motivated quark mass definition for this
purpose are the threshold masses derived in
Refs.~\cite{Erler:2004in,Erler:2017knj}. In our calculation we use these
quark masses in all places where mass-dependent terms remain after expanding in large $m_\text{weak}^2$.  However, for consistency, we exclude two-loop self-energy diagrams involving only quark
and photon propagators in the loops, such as Fig.~\ref{fig:diaga}~(a) with a
photon inside the loop, since QED effects are already subsumed in the
non-perturbative hadron dynamics.
In addition, following Ref.~\cite{Czarnecki:1995fw}, we also set
$Q^2\to 0$ in the $t$- and $u$-channel self-energies, since the differences
$\Pi_{\rm T}^{\gamma\gamma}(t)-\Pi_{\rm T}^{\gamma\gamma}(0)$ and $\Pi_{\rm
T}^{\gamma Z}(t)-\Pi_{\rm T}^{\gamma Z}(0)$ are estimated to be negligibly small
\cite{Czarnecki:1995fw}  (similar for $t$ replaced by $u$). We leave a more
detailed study of hadronic effects for future work.

As shown in Eq.~\eqref{eq:alr}, $A_{LR}$ is commonly normalized in terms of the
Fermi constant $G_\mu$, which is related to SM parameters according to
\begin{equation}
\frac{G_\mu}{\sqrt{2}} = \frac{\pi\alpha}{2\sw^2\cw^2m_Z^2}(1+\Delta r),
\end{equation}
where $\Delta r$ includes radiative corrections. The required two-loop
contributions to $\Delta r$ with one or two closed fermion loops have been taken
from Refs.~\cite{Freitas:2000gg,Freitas:2002ja} (see also Ref.~\cite{Awramik:2003ee}).

The calculation has been carried out with extensive use of computer algebra
tools. Diagrams and amplitudes were generated with {\sc FeynArts}
\cite{Hahn:2000kx}. For the Lorentz and Dirac algebra, we employed {\sc
Package-X} \cite{Patel:2015tea} and cross-checked against private code written
in {\sc Mathematica}. The large-mass expansion was implemented in-house in two
independently developed {\sc Mathematica} programs. Two-loop integrals with
non-trivial numerator structures have been reduced to simple scalar integrals
using {\sc FIRE 5} \cite{Smirnov:2014hma} and using private code based on
Ref.~\cite{Weiglein:1993hd,Davydychev:1995nq}.  For basic one-loop integrals and
two-loop vacuum integrals, analytical formulae are available
\cite{tHooft:1978jhc,Beenakker:1988jr,Ford:1992pn,Davydychev:1992mt}. We have
numerically checked our implementation of the one-loop formulae against the {\sc
Collier} library \cite{Denner:2016kdg}. The numerical dispersion integrals for
two-loop vertex and box integrals have been implemented in C and {\sc
Mathematica}. 

Each building block of the final result has been computed in two independent
setups within our collaboration and cross-checked against each other. We have
confirmed cancellation of UV and IR divergencies in the full result by
verifying that the coefficients of the $1/\epsilon$, $\log m_e^2$ and $\log
m_\gamma^2$ terms vanish algebraically. Furthermore, as an intermediate step, 
we have reproduced the one-loop result of Ref.~\cite{Czarnecki:1995fw} and
found exact agreement with the analytical formulae given there.

\section{Results}

To evaluate the numerical impact of the closed fermion-loop NNLO corrections to $A_{LR}$, we used the following input parameters:
\begin{align}
&\begin{aligned}
m_Z &= 91.1876\,\text{GeV}, & \sw^2 &= 0.2314, \\
m_H &= 125.1\,\text{GeV}, & m_t &= 173.0\,\text{GeV}, \\
m_\tau &= 1.777\,\text{GeV}, & m_b &= 3.99\,\text{GeV}, \\
m_\mu &= 105.7\,\text{MeV}, & m_c &= 1.185\,\text{GeV}, \\
m_e &= 0.511\,\text{MeV}, & m_s &= 0.342^{+0.048}_{-0.053}\,\text{GeV}, \\
&& m_{u,d} &= 0.246^{+0.054}_{-0.057}\,\text{GeV}, 
\end{aligned} \notag \\
&\Delta\alpha = 0.02761_\text{had.}+0.0314976_\text{lep.}, \label{eq:input1}
\end{align}
at the representative kinematic point
\begin{align}
&\sqrt{s} = 11\,\text{MeV}, \qquad y = 0.4.  \label{eq:input2}
\end{align}
Here $\Delta\alpha$ accounts for the renormalization group running of the fine structure constant between scales $\mu=0$ and $\mu=m_Z$, and enters our calculation through the OS charge renormalization.  The first number reflects the hadronic contribution to $\Delta\alpha$, which is obtained from $e^+e^- \to {\rm hadrons}$ data (see Refs.~\cite{Jegerlehner:2017zsb,Davier:2019can,Keshavarzi:2019abf} for recent evaluations), while the second number is the perturbatively calculable leptonic contribution \cite{Steinhauser:1998rq,Sturm:2013uka}.

As explained above, the light fermion masses $m_f,\,f\neq t$ enter in loop integrals with a fermionic photon or $\gamma$-$Z$ self-energy subloop. The values for the light quark masses are taken from Ref.~\cite{Erler:2017knj}. There is a strong anti-correlation between the reported uncertainties of $m_s$ and $m_{u,d}$. We will assume them to be 100\% anti-correlated for the results which we present below.

With these inputs we obtain numerical results for the asymmetry (\ref{eq:alr}) as shown in Table \ref{tab:results}.  The first row corresponds to the tree-level contribution, and the remaining rows $\Delta Q^e_{W(L,n_f)}$ are the radiative corrections with $L$ loops and $n_f$ closed fermion loops. No resummation of logarithms has been carried out.  In particular, the electroweak logarithms, which conventionally define the running $\sin^2\theta_W$, are left explicitly in the one and two loop results.  The last two rows $\Delta Q^e_{W(2,2)}$ and $\Delta Q^e_{W(2,1)}$ are obtained using our newly computed NNLO corrections to the asymmetry.  The error intervals reflect the hadronic uncertainties due to the threshold quark masses in Eq.~\eqref{eq:input1}.

\begin{table}
\begin{ruledtabular}
\begin{tabular}{lll}
Quantity & Contribution ($\times 10^{-3}$) \\
\hline
$1-4\sin^2\theta_W$ & $+74.4$ \\ 
\hline
	$\Delta Q^e_{W(1,1)}$ & $-29.0$ \\ 
	$\Delta Q^e_{W(1,0)}$ & $+\phantom{0}3.1$ \\ 
\hline
	$\Delta Q^e_{W(2,2)}$ & $-\phantom{0}0.18^{+0.0024}_{-0.0040}$ \\ 
	$\Delta Q^e_{W(2,1)}$ & $+\phantom{0}1.18^{+0.015}_{-0.010}$ \\ 
	$\Delta Q^e_{W(2,0)}$ & $\pm\phantom{0}0.13$ (estimate) \\ 
\end{tabular}
\end{ruledtabular}
\caption{Numerical estimates of the calculated contributions to the polarized M{\o}ller scattering asymmetry defined in (\ref{eq:alr}) through NNLO using input values in (\ref{eq:input1}) and (\ref{eq:input2}). Subscripted indices on $\Delta Q^e_{W(L,n_f)}$ refer to the loop order $L$ and number of closed loops $n_f$.}
\label{tab:results}
\end{table}

The precision goal for the MOLLER experiment corresponds to a measurement of the weak charge with an uncertainty of $\delta_{\rm exp}Q^e_W = 1.1\times 10^{-3}$. The NNLO corrections with closed fermion loops add up to
\begin{equation}\label{eq:tot}
\Delta Q^e_{W(2,2)} + \Delta Q^e_{W(2,1)} = 1.00^{+0.012}_{-0.008} \times 10^{-3}\,,
\end{equation}
which is comparable to the experimental target, thus highlighting the importance of accounting for the NNLO corrections. On the other hand, anti-correlation between the  hadronic uncertainties in the individual contributions with one and two closed fermion loops leads to a reduced overall hadronic uncertainty.

The resulting hadronic uncertainty from quark loops is negligible compared to the experimental target precision. It is likely that the our estimate based on quark mass errors overestimates this uncertainty, since we cannot account for correlations between the quark masses and the $K$ factors in Ref.~\cite{Erler:2017knj}\footnote{In fact, when estimating the leading hadronic effects by plugging these quark masses into the NLO correction, one finds an uncertainty that is a factor few larger than the detailed renormalization-group evaluation in Ref.~\cite{Erler:2017knj}.}. A more detailed analysis of hadronic effects will be given in a future publication.

The correction $\Delta Q^e_W$ depends very mildly on $y$ (i.e.\ on the scattering angle). Varying $y$ in the  experimentally relevant range $(0.25,0.75)$ \cite{Benesch:2014bas}, we find that $\Delta Q^e_W$ changes by $0.04 \times 10^{-3}$ for the NLO corrections, and by $0.01\times 10^{-3}$ for the NNLO corrections, both of which are negligible.

Finally, we attempt to estimate size of the currently missing NNLO corrections without closed fermion loops $\Delta Q^e_{W(2,0)}$ (called ``bosonic'' corrections in the following). For this purpose, we begin by comparing the relative size of the fermion-loop and bosonic correction at NLO.  From Table \ref{tab:results}, these are $\Delta Q^e_{W(1,1)} = -0.0290$ and $\Delta Q^e_{W(1,0)}=+0.0031$, respectively. Assuming a similar ratio between the corrections with one closed fermion loop and the bosonic corrections at NNLO, we obtain an estimate of $0.13\times 10^{-3}$ for the size of the latter. This would be safely below the experimental target precision.

\section{Conclusions}

To correctly interpret the proposed 2.4\% measurement of the parity-violating asymmetry $A_{LR}$ from the MOLLER experiment at the Jefferson Lab in terms of BSM physics, we calculate the NNLO SM contributions to $A_{LR}$ using large-mass expansion and numerical integration of sub-loop dispersion relation.  We summarize our results in Table~\ref{tab:results}. 
We find that the corrections to $\Delta Q_W^e$ from diagrams with closed fermion loops are comparable to the experimental target precision.
The dependence of  $\Delta Q_W^e$ on the scattering angle is very mild in the experimentally relevant range and can be ignored for most practical purposes. Finally, we also consider the impact of the remaining bosonic NNLO corrections and estimate them to be negligible compared to the MOLLER precision goal.  However, it is desirable to confirm this with an explicit calculation in the future.

\section*{Acknowledgments}

The authors would like to thank V.A.~Smirnov and J.~Erler for useful private communications. YD is grateful to 
K.~Kumar for his financial support at Fermilab and to PITT PACC at the University of Pittsburgh for their hospitality during part of this work.
A.F.\ has been supported in part by the National Science Foundation under
grant no.\ PHY-1820760. Y.D., H.H.P.\ and MJRM were supported in part under U.S.\ Department of Energy contract DE-SC0011095.  H.H.P. was additionally supported in part by U.S. Department of Energy grant number de-sc0010107.

\bibliographystyle{utcaps_mod}
\bibliography{mollerbib}

\end{document}